\documentclass[11pt]{article}
\usepackage[brazil]{babel}
\usepackage{url}
\usepackage{amssymb,amsxtra}
\usepackage{amsmath,color,graphicx}
\usepackage{lineno}
\usepackage{graphicx}
\RequirePackage{geometry}
\usepackage{indentfirst}
\usepackage[breakable]{tcolorbox}
\usepackage{lineno,hyperref}
\usepackage{alltt}
\usepackage{amsfonts}
\usepackage{amsmath,amssymb}
\usepackage{multirow}
\usepackage{eqparbox}
\usepackage{upquote} 
    \usepackage{eurosym} 
    \usepackage[mathletters]{ucs} 
    \usepackage{fancyvrb} 
    \usepackage{grffile} 
\definecolor{incolor}{HTML}{303F9F}
    \definecolor{outcolor}{HTML}{D84315}
    \definecolor{cellborder}{HTML}{CFCFCF}
\definecolor{cellbackground}{HTML}{F7F7F7}
\definecolor{darkgreen}{rgb}{0.00,0.50,0.00}
\definecolor{purple}{rgb}{0.67,0.13,1.00}
\definecolor{lgray}{rgb}{0.60,0.60,0.60}

\begin{document}

\def\chaptername{}
\def\contentsname{Sum\'{a}rio}
\def\listfigurename{Figuras}
\def\listtablename{Tabelas}
\def\abstractname{Resumo}
\def\appendixname{Ap\^{e}ndice}
\def\refname{\large Refer\^{e}ncias bibliogr\'{a}ficas}
\def\bibname{Bibliografia}
\def\indexname{\'{I}ndice remissivo}
\def\figurename{\small Fig.~}
\def\tablename{\small Tab.~}
\def\pagename{\small Pag.}
\def\seename{veja}
\def\alsoname{veja tamb\'em}
\def\na{-\kern-.4em\raise.8ex\hbox{{\tt \scriptsize a}}\ }
\def\pa{\slash \kern-.5em\raise.1ex\hbox{p}\ }
\def\ro{-\kern-.4em\raise.8ex\hbox{{\tt \scriptsize o}}\ }
\def\no{n$^{\underline{\rm o}}$}

\setcounter{tocdepth}{3}

\clearpage
\pagenumbering{arabic}

\thispagestyle{empty}
\parskip 8pt

\vspace*{0.2cm}
\begin{center}

{\huge \bf Aplica\c{c}\~{o}es do m\'{e}todo de Numerov a sistemas qu\^{a}nticos simples usando Python}\\

\ \\

{\huge \bf  Applications of the Numerov method to simple quantum systems using Python}\\

\vspace*{1.0cm}
{\Large \bf \it Francisco Caruso;\,$^{1,2}$ Vitor Oguri;\,$^{2}$ Felipe Silveira\,$^{2}$}\\[2.em]

{{$^{1}$ Centro Brasileiro de Pesquisas F\'{\i}sicas, Coordena\c{c}\~{a}o de F\'{\i}sica de Altas Energias, 22290-180, Rio de Janeiro, RJ, Brasil.}}

{{$^{2}$ Universidade do Estado do Rio de Janeiro, Instituto de F\'{\i}sica Armando Dias Tavares, 20550-900, Rio de Janeiro, RJ, Brasil.}}
\vfill
\end{center}

\noindent \textbf{Resumo}

O m\'{e}todo num\'{e}rico de Numerov \'{e} desenvolvido de forma did\'{a}tica usando Python no {\it Jupyter Notebook} vers\~{a}o 6.0.3 para tr\^{e}s diferentes sistemas da f\'{\i}sica qu\^{a}ntica: o \'{a}tomo de hidrog\^{e}nio, uma mol\'{e}cula governada pelo potencial de Morse e um {\it quantum dot}. Ap\'{o}s uma breve introdu\c{c}\~{a}o ao m\'{e}todo Numerov, \'{e} apresentado o c\'{o}digo completo para calcular as autofun\c{c}\~{o}es e autovalores do \'{a}tomo de hidrog\^{e}nio. As altera\c{c}\~{o}es de c\'{o}digo necess\'{a}rias para calcular os outros dois exemplos tamb\'{e}m s\~{a}o fornecidas em sequ\^{e}ncia.

\noindent \textbf{Palavras-chave:} \'{a}tomo de hidrog\^{e}nio; potencial de Morse; Quantum Dot; metodo de Numerov; Python.

\vspace*{0.7cm}

\noindent \textbf{Abstract}

Numerov's numerical method is developed in a didactic way by using Python in its {\it Jupyter Notebook} version 6.0.3 for three different quantum physical systems: the hydrogen atom, a molecule governed by the Morse potential and for a quantum dot. After a brief introduction to the Numerov method,
the complete code to calculate the eigenfunctions and eigenvalues of the hydrogen atom is presented. The necessary code changes to calculate the other two examples are also provided in the sequel.


\vfill

\newpage

\section{Introduction} \label{intro}

The vast majority of numerical methods, such those of Newton, Euler, Lagrange, Gauss, Fourier, Jacobi, Runge-Kutta and so many others~\cite{Numerov6}, were introduced in the context of applications in physics, astronomy or in other areas of a technical nature, such as aerodynamics. Since then, numerical analysis was not being recognized as a mathematical discipline and this situation persisted during the first four decades of 20th century. Even today, although some numerical methods are taught in physics courses, within the disciplines of mathematics, little emphasis is given to them in physical applications.

In the las decades, the teaching of computing techniques has become more present and also increasingly essential in the development of students from all areas, and therefore, it would not be different for physics teaching. Having said that, it is of paramount importance that we always produce new teaching materials for new technologies such as the Python programming language, which, despite of being relatively new, has already dominated the market to become one of the most important languages today.

We will disclose here a powerful numerical calculation method originally developed by Boris Vasil’evich Numerov~\cite{Numerov, Numerov2}, see also~\cite{Numerov3, Numerov4, Numerov5, baugci2021efficient}, applying it to the time-independent Schr\"{o}dinger equation describing physical systems like the hydrogen atom, a diatomic molecule governed by the Morse potential and one model for the quantum dot atom~\cite{Caruso, caruso2017corrigendum, BJP}. These three examples will be solved and the parameters needed for each solution using Numerov's method will be shared in their respective sections.

In summary, this work aims to provide the complete code developed in Python with the {\it Jupyter Notebook} for the Numerov's numerical method. However, it is important to emphasize that we do not aim to teach Python to the reader, who must have a basic knowledge of programming to be able to keep up the examples.

\section{Numerov's Method} \label{NM}
Numerov's initial motivation was to be able to calculate corrections to the trajectory of comet Halley. Therefore, Numerov's method was initially developed to determine solutions to eigenvalue problems associated with ordinary differential equations of second order of celestial mechanics, which did not contain terms involving the first derivative of a function unknown $y(x)$, that is, equations of the form

\begin{equation}\label{dif}
  \frac{\mbox{d}^2y}{\mbox{d}x^2} = f(y,x).
\end{equation}

Every differential equation equal to equation~(\ref{dif}) can be replaced by the following system of first order equations

\begin{equation*}
  \left\{
                \begin{array}{ll}
                 \displaystyle \frac{\mbox{d}z}{\mbox{d}x} = f(x,y),\\
                 \displaystyle z = \frac{\mbox{d}y}{\mbox{d}x}.\\
                \end{array}
              \right.
\end{equation*}

Traditional methods for numerically solving this system of equations, such as those of Euler or Runge-Kutta, consider that the values of $y(x)$ and of $\mbox{d}y/\mbox{d}x$ are known at a given point in the domain $[a, b]$ of system validity, \textit{i.e.}, are suitable for the so-called seed problems.

In non-relativistic quantum mechanics, more specifically in bound state problems involving a particle of mass $m$ confined in a well of potential $V(x)$, in a given interval $a < x < b$, the allowed energies $(E)$ and the corresponding wave functions $\psi(x)$ that describe these steady states satisfy the Schr\"{o}dinger's eigenvalue equation

\begin{equation}\label{sch1}
  \frac{\mbox{d}^2\psi}{\mbox{d}x^2} + k^2(x) \psi = 0
\end{equation}
where $k = \sqrt{2m[E-V(x)]}/\hbar$ and $\hbar \simeq 1.055 \times 10^{-34} J.s$ is the reduced Planck constant.

In these cases, as the value of the first derivative of the wave function is not known, the Euler and Runge-Kutta cannot be employed. Nonetheless, it is possible to establish continuity conditions for the values of $\psi$ and $\mbox{d}\psi/\mbox{d}x$ at two or more points of the domain of the wave function, which characterizes the so-called boundary value problems.

In addition to making the transformation of a second order differential equation in a first order system, the Numerov method allows the simultaneous determination of the energy spectrum of the particle and of the eigenfunctions associated with each energy value.

Like any iterative numerical method, the solution of equation~(\ref{sch1}) is constructed by successive integrations. In Numerov's method, initially, the solution is considered to be known at two subsequent points of the interval $[a, b]$, for example, at $\psi(x - \delta)$ and $\psi(x)$, where $\delta$ is an arbitrarily small quantity, called the integration step. Next, we try to establish an algorithm to determine the solution at the next point, $\psi(x + \delta)$.

The starting point for establishing this algorithm is the expansion of $\psi(x \pm \delta)$ in Taylor series, up to fourth-order derivatives, that is,

\begin{equation}\label{taylor}
  \psi(x\pm\delta)=\psi(x) \pm \delta \psi'(x) + \frac{\delta^2}{2}\psi''(x) \pm \frac{\delta^3}{6}\psi'''(x) + \frac{\delta^4}{24}\psi^{iv}(x)
\end{equation}

Adding the terms $\psi(x + \delta)$ and $\psi( x - \delta)$, only the derivatives of even order survive and, therefore, a relationship between the values of a function in three is reached. points and its second derivative, given by

\begin{equation}\label{taylor2}
  \frac{\psi(x+\delta)+\psi(x-\delta)-2\psi(x)}{\delta^2} = \psi''(x)+\frac{\delta^2}{12}\psi^{iv}(x) \equiv \left(1+\frac{\delta^2}{12}\frac{\mbox{d}^2}{\mbox{d}x^2}\right)\psi''{x}
\end{equation}

Writing the unidimensional Schr\"{o}dinger equation, equation~(\ref{sch1}), in a more convenient form
\begin{equation}\label{taylor3}
\left(1 + \frac{\delta^2}{12}\frac{\mbox{d}^2}{\mbox{d}x^2}\right)\psi''(x) = -k^2(x) \psi(x) - \frac{\delta^2}{12}\frac{\mbox{d}^2}{\mbox{d}x^2}\left[k^2(x)\psi(x)\right]
\end{equation}

\noindent and using equation~(\ref{taylor2}) to replace the terms that contain second order derivatives, one obtains

\begin{eqnarray}\label{taylor4}
   &&\frac{\psi(x+\delta) + \psi(x-\delta) - 2\psi(x)}{\delta^2} = -k^2(x)\psi(x) \\
   && - \frac{\delta^2}{12} \times \left[\frac{k^2(x+\delta)\psi(x+\delta)+k^2 (x-\delta)\psi(x-\delta) - 2k^2(x)\psi(x)}{\delta^2}\right] + \mathcal{O}(\delta^4)  \nonumber
\end{eqnarray}

\noindent Regrouping the therm we obtain the Numerov difference formula for the problem of a particle under action of a one-dimensional potential

\begin{equation}\label{taylor5}
  \left[1 + \frac{h^2}{12} k^2(x+\delta)\right]\psi(x+\delta) = 2\left[1-\frac{5\delta^2}{12}k^2(x)\right]\psi(x) - \left[1+\frac{\delta^2}{12}k^2(x-\delta)\right]\psi(x-\delta)
\end{equation}

In fact, it should be noted that the algorithm can be applied to any ordinary linear differential equation and second-order homogeneous that does not contain terms of first derivative.

Since the problem of interest is an eigenvalue problem, the numerical integration technique of the one-dimensional Schr\"{o}dinger equation for a particle in a well depends on attaching arbitrary values conveniently to eigenvalues and to the respective (possible) eigenfunctions in 2 points of the domain of the problem. But how to do it? Regarding the choice of the initial value for the energy (first eigenvalue), just remember that, according to Heisenberg uncertainty relation, the energy $E$ of a particle in a well of potential $V(x)$ must be greater than the minimum value of the well. Thus, it is considered, initially, that $E_{initial} = V_{min} + \Delta E$, with $\Delta E >0$

\begin{figure}[ht]
\centerline{\includegraphics[width=8.0cm]{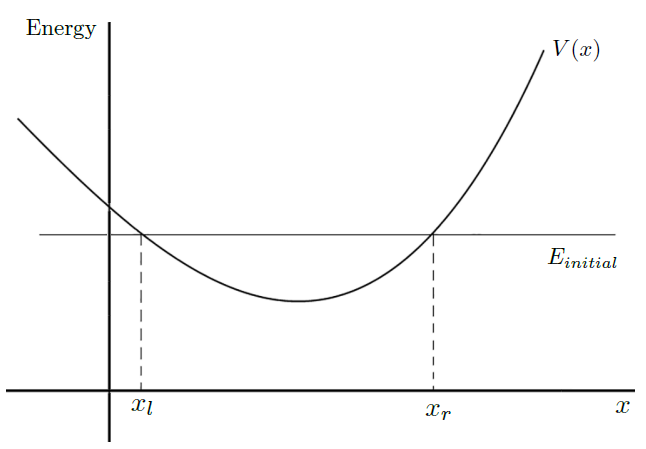}}
\caption{Meeting points between the potential curve and the initial energy, also called turning points.}
\label{fig1}
\end{figure}

The choice of a energy value, determines two turning points, $x_{\ell}$ and $x_{r}$, where the energy value is equal to potential energy value, whose motion obeys the classical Newtonian mechanics. That is, from the point of view of classical mechanics, the movement of the particle is restricted only to the region $\left[x_{\ell},x_{r}\right]$, in which the energy is greater than or equal to the potential energy. The regions $x<x_{\ell}$ and $x>x_{r}$ are called classically prohibited regions, and are indicated in Fig.~\ref{fig1}

As the Schr\"{o}dinger equation admits solutions for these classically prohibited regions, for each energy value, initially, values are assigned to a possible eigenfunction at two points of the classically prohibited regions, in which the function practically cancels itself. In general, these are the boundary points $a$ and $b>a$ of the function's integration domain.

However, the implementation of Numerov's method to solve the problem still requires an iteration scheme that uses the Numerov formula in two steps: from $a$, or to the left of $a$ from the classic turning points, hereinafter called match point ($x_{match}$), and from $b$, or to the right of the match point.

Thus, arbitrarily taking a initial value for the energy, and two successive arbitrary values for the solution, starting from the lower extremes and upper part of the integration interval $\left[a,b\right]$, one can implement the method's iteration scheme in the two senses, such as:

\begin{itemize}
  \item Solution to the left of match point $(x<x_{match})$
\end{itemize}

Being $E_{initial} = V_{min} + \Delta E \left(\Delta E/|V_{min}|\ll1\right)$ an arbitrary value for the energy of the particle. Also arbitrating values for the function of wave, in 2 successive points, from $a$,

\begin{equation*}
  \left\{
                \begin{array}{ll}
                  \psi^{\ell}(a)=0,\\
                  \psi^{\ell}(a+\delta) = \delta^\ell,\quad (\delta^\ell \ll 1)\\
                \end{array}
              \right.
\end{equation*}

\noindent and using the formula of differences, equation~(\ref{taylor5}), the solution on the left is built sequentially until match point $(x_{match})$, in what $\psi^\ell(x_{match}) = \psi^\ell_{match}$.

\begin{itemize}
  \item Solution to the right of match point $(x>x_{match})$
\end{itemize}

From a similar way, for the same values $E_{initial}$, arbitrating

\begin{equation*}
  \left\{
                \begin{array}{ll}
                  \psi^{r}(b)=0,\\
                  \psi^{r}(b+\delta) = \delta^r,\quad (\delta^r \ll 1)\\
                \end{array}
              \right.
\end{equation*}

the solution to the right, from $b$, is constituted sequentially until the points $x_{match}$ and $x = x_{match}-\delta$, as

\begin{equation*}
  \left\{
                \begin{array}{ll}
                  \psi^{r}(x_{match})=\psi^{r}_{match},\\
                  \psi^{r}(x_{match} - \delta) = \psi^r_{match -1},\quad (\delta^r \ll 1)\\
                \end{array}
              \right.
\end{equation*}

To guarantee the boundary condition of the solution, we redefine the solution to the left according to equation~(\ref{eq8}) given below, and the boundary condition of the first derivatives, according to equation~(\ref{eq9}).

The procedure is repeated step by step, in the two ways, $a \rightleftharpoons b$. Starting from $a$, using the Numerov recurrence formula associated with an equation, if we build the solution $\psi^\ell$ until the classic rewind point, nearest of $b$, where $E = V(x_{match})$, called the match point. Then, from $b$, the analog is made, building a solution $\psi^r$ to the match point. In principle, the possible solutions $\psi^\ell$ and $\psi^r$ will not necessarily be equal in this $x_{match}$ stitch. To ensure the continuity of the solution redefines itself $\psi^\ell$ like

\begin{equation}\label{eq8}
  \psi^\ell(x) \rightarrow \psi^\ell(x) \frac{\psi^r (x_{matxh})}{\psi^\ell (x_{matxh})} \quad (a \leq x \leq x_{match})
\end{equation}

Finally, it is verified how close are the values of the respective first derivatives of $\psi^r$ and the new function $\psi^\ell$ It is staggered, at match point. To test the boundary condition of the derivatives first, taking into account Taylor's series for $\psi(x+\delta$ and $\psi(x-\delta)$, up to the first order, you can write

\begin{equation}\label{eq9}
   \left\{
                \begin{array}{ll}
                  \frac{\mbox{d}\psi^\ell}{\mbox{d}x} \big{|}_{x_r} = \frac{\psi^\ell_{match+1}-\psi^\ell_{match-1}}{2\delta},\\
                  \frac{\mbox{d}\psi^r}{\mbox{d}x} \big{|}_{x_r} = \frac{\psi^r_{match+1}-\psi^r_{match-1}}{2\delta},\\
                \end{array}
              \right.
\end{equation}

\noindent in what, $\psi_{match\pm1} = \psi(x_{match\pm\delta})$.

If the difference between these values is less than the values of a predefined error, the process is interrupted, confirming the searched eigenvalue and the respective eigenfunction as being

\begin{equation*}
  \left\{
                \begin{array}{ll}
                  \psi^{\ell}(x), \quad (a \leq x < x_{match})\\
                  \psi^{r}(x), \quad (x_{match} \leq x \leq b)\\
                \end{array}
              \right.
\end{equation*}

If the continuity condition of the derivatives is not satisfied, the value of the energy is increased and we restarted a search for a new value which is really an eigenvalue of the problem, and its respective eigenfunction.
The process can be repeated until the desired number of eigenvalues and eigenfunctions of the problem.

Because it is based on Taylor's serial expansion to fourth order, the error in Numerov's method is much smaller than the errors that come out from the expansion-based methods in lower order, like that of Runge-Kutta.
\newpage
\section{Hidrogen Atom} \label{atom}
Numerical solutions of hydrogen atom was previously obtained in~\cite{Numerov6} using Numerov's method. The program was written in C++ for the ROOT cint compiler.

Although it was originally developed for second order linear and homogeneous ordinary differential equations that do not contain terms of the first derivative, the
Numerov's method can be generalized to cover the presence of terms that contain the first derivative in the differential equation, so that eigenvalue problems can also be considered.

In fact, in the case of linear equations, every equation second order differential of type

\begin{equation*}
  \frac{\mbox{d}^2y}{\mbox{d}x^2} + P(x) \frac{\mbox{d}y}{\mbox{d}x} + Q(x)y = 0
\end{equation*}

\noindent can be written in its normal form

\begin{equation*}
  \frac{\mbox{d}^2y}{\mbox{d}x^2} = q(x)y = 0
\end{equation*}

\noindent where

\begin{equation*}
  q(x) = Q(x) - \frac{1}{4} P^2(x) - \frac{1}{2}\frac{\mbox{d}P}{\mbox{d}x}
\end{equation*}

Schr\"{o}dinger's radial equation for a particle of mass $m$ under the action of a Coulombian electric field, like the electron in the hydrogen atom, can be
written as
\begin{equation}\label{sch}
  \frac{\mbox{d}^2R(r)}{\mbox{d}r^2} + \frac{2}{r} \frac{\mbox{d}R(r)}{\mbox{d}r} + \frac{2m}{\hbar^2}\left[E+\frac{e^2}{r} - \frac{\hbar^2}{2m} \frac{\ell(\ell+1)}{r^2}\right]R(r) = 0
\end{equation}

Making the substitution $r = xa_B$ with $a_B = \hbar^2/(me^2)$ being the Bohr radius, equation~(\ref{sch}) can be rewritten, for a new function $y(x) = R(r)$ as

\begin{equation}\label{sch2}
  \frac{\mbox{d}^2y}{\mbox{d}x^2} = -\frac{2}{x}\frac{\mbox{d}y}{\mbox{d}x} - \left[\varepsilon - V(x)\right]y(x)
\end{equation}

\noindent where, $\varepsilon = \frac{E}{e^2/(2a_B)}$ and $V(x) = \frac{\ell(\ell+1)}{x^2} - \frac{2}{x}$ are, respectively, the energy and the so-called effective potential in atomic units. So, in possession of equation~(\ref{sch2}), we can start building our program code.

First of all, we must import the functions available in the Pylab module that bulk imports matplotlib.pyplot (for plotting) and NumPy (for Mathematics and working with arrays) in a single name space. We then declare who our effective potential $V(x)$ is, and during the construction of this example we will use $\ell=0$.

\vspace*{0.1cm}
\begin{tcolorbox}[breakable, size=fbox, boxrule=1pt, pad at break*=1mm,colback=cellbackground, colframe=cellborder]
\begin{Verbatim}[commandchars=\\\{\}, baselinestretch=1.0]
 {\color{darkgreen}{\bf from}} pylab {\color{darkgreen}{\bf import}} {\color{purple}*}
 x {\color{purple}=} linspace({\color{purple}-}{\color{darkgreen}10},{\color{darkgreen}10},{\color{darkgreen}1001})
 {\color{darkgreen}{\bf def}} V(x):
    L{\color{purple}=}{\color{darkgreen}0}
    Vx{\color{purple}=}{\color{purple}-}{\color{darkgreen}2}.{\color{purple}{\color{purple}/}}x
    {\color{darkgreen}{\bf return}}  Vx
\end{Verbatim}
\end{tcolorbox}
\vspace*{0.1cm}

\vspace*{0.1cm}
\begin{tcolorbox}[breakable, size=fbox, boxrule=1pt, pad at break*=1mm,colback=cellbackground, colframe=cellborder]
\begin{Verbatim}[commandchars=\\\{\}, baselinestretch=1.0]
{\color{darkgreen}def} vec_max(dim,x):                #left maximum
    xmax{\color{purple}=}{\color{darkgreen}0}
    N{\color{purple}=}dim
    {\color{darkgreen}for} j in range(int(N)):
        if j<int(N) and abs(x[j])>xmax:
               xmax{\color{purple}=}abs(x[j])
        else:
               continue
    {\color{darkgreen}return} xmax
\end{Verbatim}
\end{tcolorbox}
\vspace*{0.1cm}

The equation, in this case, that is intended to be solved by Numerov's method presents a term involving the first derivative, and can be expressed by

\begin{equation}\label{13}
  \psi''(x) = -p(x)\psi'(x) - s (x)\psi(x),
\end{equation}

where

\begin{equation*}
  \left\{
               \displaystyle \begin{array}{ll}
               \displaystyle   p(x)=\frac{2}{x} \quad \Rightarrow \quad p'(x) = - \frac{2}{x^2},\\
               \displaystyle   s(x) = \varepsilon - V(x).\\
                \end{array}
              \right.
\end{equation*}

From the Taylor expansions, equation~(\ref{taylor}), we can rewrite equation~(\ref{13}) as

\begin{equation}\label{14}
  \left(1+\frac{\delta^2}{12}\frac{\mbox{d}^2}{\mbox{d}x^2}\right)\psi''(x) = -p(x)\psi'(x) - s(x)\psi(x) - \frac{\delta^2}{12}\frac{\mbox{d}^2}{\mbox{d}x^2}\left[p(x)\psi'(x) + s(x)\psi(x)\right].
\end{equation}

In a similar way to the previous case, according to equation~(\ref{taylor2}), you can write the term on the right side of the equation~(\ref{14}) which contains derivatives of order 2, such as

\begin{eqnarray*}
  &&\frac{\mbox{d}^2}{\mbox{d}x^2} \left[p(x)\psi'(x) + s(x)\psi(x)\right] = \frac{1}{\delta^2} \left[p(x+\delta)\psi'(x+\delta) + s(x+\delta)\psi(x+\delta) +\right. \\
  && + \left.  p(x-\delta)\psi'(x-\delta) + s(x-\delta)\psi(x-\delta) + -2p(x)\psi'(x)-2s(x)\psi(x)\right]
\end{eqnarray*}

Replacing first order derivatives with approximations

\begin{equation*}
  \left\{
                \begin{array}{lll}
                  \psi'(x) = \left[\psi(x+\delta) - \psi(x-\delta)\right]/(2\delta),\\
                  \psi'(x+\delta) = \left[\psi(x+\delta) - \psi(x)\right]/(\delta),\\
                  \psi'(x-\delta) = \left[\psi(x) - \psi(x-\delta)\right]/(\delta),\\
                \end{array}
              \right.
\end{equation*}

\noindent we obtain

\begin{eqnarray*}
&& \frac{\mbox{d}^2}{\mbox{d}x^2}\left[p(x)\psi'(x) + s(x)\psi(x)\right] = \frac{1}{\delta^2} \left\{ \left[\frac{p(x+\delta)-p(x)}{\delta}+s(x+\delta)\right]\times\right.  \\
&&  \times \psi(x+\delta)+ \left.\left[\frac{p(x)-p(x-\delta)}{\delta}+s(x-\delta)\right]\psi(x-\delta) +\right. \\
&& + \left.2\left[\frac{p(x-\delta)-p(x+\delta)}{2\delta}+s(x)\right]\psi(x)  \right\}
\end{eqnarray*}

\noindent or

\begin{eqnarray}\label{17}
   && \frac{\mbox{d}^2}{\mbox{d}x^2} \left[p(x)\psi'(x) + s(x)\psi(x)\right] = \frac{1}{\delta^2} \left\{ \left[p'(x)+s(x+\delta)\right]\psi(x+\delta)+\right.  \nonumber\\
   && + \left.\left[p'(x)+s(x-\delta)\right]\psi(x-\delta) - 2\left[p'(x)+s(x)\right]\psi(x)  \right\}
\end{eqnarray}

Taking into account that the left side of the equation~(\ref{14}) is equal to

\begin{equation*}
  \left[\psi(x+\delta)+\psi(x-\delta)-2\psi(x)\right]/\delta^2
\end{equation*}

\noindent we can write

\begin{eqnarray*}
 &&\frac{\psi(x+\delta)+\psi(x-\delta) - 2\psi(x)}{\delta^2} = -p(x)\left[\frac{\psi(x+\delta)-\psi(x-\delta)}{2\delta}\right] \\
  && - s(x)\psi(x) + \frac{1}{12}\left[p'(x)+s(x+\delta)\right]\psi(x+\delta) \\ &&
   - \frac{1}{12}\left[p'(x)+s(x-\delta)\right]\psi(x-\delta) + \frac{1}{6}\left[p'(x)+s(x)\right]\psi(x)
\end{eqnarray*}

Regrouping the terms, and making

\begin{equation*}
  \left\{
                \begin{array}{lll}
                  \psi(x-\delta)=\psi_0,\\
                  r\psi(x)=\psi_1,\\
                  \psi(x+\delta) = \psi_2,\\
                \end{array}
              \right.
\end{equation*}

\noindent one obtains the Numerov difference equation for the problem, suitable for the propagation of the solution from of the limits of the integration interval

\begin{equation}\label{19}
  \psi_2 = \frac{2\left\{1-\left[s(X)-\frac{p'(x)}{5}\right]\frac{5\delta^2}{12}\right\}\psi_1 -\left\{1-p(x) \frac{\delta}{2} + \left[s(x-\delta)+p'(x)\right]\frac{\delta^2}{12} \right\}\psi_0}{\left\{1+p(x)\frac{\delta}{2}+\left[s(x+\delta)+p'(x)\right]\frac{\delta}{12}\right\}}
\end{equation}

From this formula, a procedure analogous to the previous case can be implemented for the construction of solutions of the radial Schr\"{o}dinger equation in the interval $(0,\infty)$.

Now, in order to introduce the Numerov diference formula~(\ref{19}), we first need to insert equation~(\ref{taylor5}) in our code, for that,  let's break it down into different pieces $p_0$, $p_1$ and $p_2$, with $h=\delta$, $q_0 = s(x)$. Thus, equation~(\ref{19}) is now called $y_2$, where $\psi_i = y_i$.

\vspace*{0.1cm}
\begin{tcolorbox}[breakable, size=fbox, boxrule=1pt, pad at break*=1mm,colback=cellbackground, colframe=cellborder]
\begin{Verbatim}[commandchars=\\\{\}, baselinestretch=1.0]
{\color{darkgreen}def} nrovl(y0, y1, x0, E, h, iflag):
    q0  {\color{purple}=}  (E{\color{purple}-}V(x0))
    q1  {\color{purple}=}  (E{\color{purple}-}V(x0{\color{purple}+}h))
    q2  {\color{purple}=}  (E{\color{purple}-}V(x0{\color{purple}+}h{\color{purple}+}h))
    p0  {\color{purple}=}  ({\color{darkgreen}1} {\color{purple}+} h{\color{purple}*}h{\color{purple}*}q0{\color{purple} /}{\color{darkgreen}12})
    p1   {\color{purple}=}   {\color{darkgreen}2}{\color{purple}*}({\color{darkgreen}1} {\color{purple}-} {\color{darkgreen}5}{\color{purple}*}h{\color{purple}*}h{\color{purple}*}q1{\color{purple} /}{\color{darkgreen}12})
    p2   {\color{purple}=}   {\color{darkgreen}1} {\color{purple}+} h{\color{purple}*}h{\color{purple}*}q2{\color{purple} /}{\color{darkgreen}12}
    y2   {\color{purple}=}  (p1{\color{purple}*}y1{\color{purple}-}p0{\color{purple}*}y0){\color{purple} /}p2
    if iflag<{\color{darkgreen}1}:
        print(" x0 {\color{purple}=} ", x0," y0 {\color{purple}=} ", y0," V {\color{purple}=} ",V(x0))
        print(" x1 {\color{purple}=} ", x0{\color{purple}+}h," y1 {\color{purple}=} ",y1," V {\color{purple}=} ",V(x0{\color{purple}+}h),
        " y2 {\color{purple}=} ",y2)
    {\color{darkgreen}return}  y2
\end{Verbatim}
\end{tcolorbox}
\vspace*{0.1cm}

And, for the case $x-\delta$, we repeat the previous step, changing the appropriate sign.

\vspace*{0.1cm}
\begin{tcolorbox}[breakable, size=fbox, boxrule=1pt, pad at break*=1mm,colback=cellbackground, colframe=cellborder]
\begin{Verbatim}[commandchars=\\\{\}, baselinestretch=1.0]
{\color{darkgreen}def} nrovr(y0, y1, x0, E, h, iflag):
    q0  {\color{purple}=}  (E{\color{purple}-}V(x0))
    q1  {\color{purple}=}  (E{\color{purple}-}V(x0{\color{purple}-}h))
    q2  {\color{purple}=}  (E{\color{purple}-}V(x0{\color{purple}-}h{\color{purple}-}h))
    p0  {\color{purple}=}  ({\color{darkgreen}1} {\color{purple}+} h{\color{purple}*}h{\color{purple}*}q0{\color{purple} /}{\color{darkgreen}12})
    p1   {\color{purple}=}  {\color{darkgreen}2}{\color{purple}*}({\color{darkgreen}1} {\color{purple}-} {\color{darkgreen}5}{\color{purple}*}h{\color{purple}*}h{\color{purple}*}q1{\color{purple} /}{\color{darkgreen}12})
    p2   {\color{purple}=}   {\color{darkgreen}1} {\color{purple}+} h{\color{purple}*}h{\color{purple}*}q2{\color{purple} /}{\color{darkgreen}12}
    y2   {\color{purple}=}  (p1{\color{purple}*}y1{\color{purple}-}p0{\color{purple}*}y0){\color{purple} /}p2
    if iflag<{\color{darkgreen}1}:
        print(" x_100 {\color{purple}=} ", x0," y0_100 {\color{purple}=} ", y0," V {\color{purple}=} ",V(x0))
        print(" x_99 {\color{purple}=} ", x0{\color{purple}-}h," y_99 {\color{purple}=} ",y1," V {\color{purple}=} ",V(x0{\color{purple}-}h),
        " y_98 {\color{purple}=} ",y2)
    {\color{darkgreen}return}  y2

{\color{darkgreen}def} espectro(xl,xu,h,delta,eps,dim,nmax,kmax,Ein,Vmax,dE,iflag):
\end{Verbatim}
\end{tcolorbox}
\vspace*{0.1cm}

We now create a list for each variable up to the value of {\it dim} or {\it nmax} which will also be defined in a future step.

\vspace*{0.1cm}
\begin{tcolorbox}[breakable, size=fbox, boxrule=1pt, pad at break*=1mm,colback=cellbackground, colframe=cellborder]
\begin{Verbatim}[commandchars=\\\{\}, baselinestretch=1.0]
    xx{\color{purple}=}list(range(dim));     yy{\color{purple}=}list(range(dim))
    ww{\color{purple}=}list(range(dim));     yl{\color{purple}=}list(range(dim))
    yr{\color{purple}=}list(range(dim));     ee{\color{purple}=}list(range(nmax))
    ff{\color{purple}=}list(range(nmax));    ff2{\color{purple}=}list(range(nmax))
    yy1{\color{purple}=}list(range(dim));    yy2{\color{purple}=}list(range(dim))
    yy3{\color{purple}=}list(range(dim))
    colors {\color{purple}=}['b','r','g','m','c'];  nk{\color{purple}=}list(range(nmax))
    E_old {\color{purple}=} Ein;   E {\color{purple}=} Ein {\color{purple}+} dE
\end{Verbatim}
\end{tcolorbox}
\vspace*{0.1cm}

In the next steps, the program will determine the interaction for the eigenvalue candidates and determine their solutions as well as printing the parameters found on the screen.

\vspace*{0.1cm}
\begin{tcolorbox}[breakable, size=fbox, boxrule=1pt, pad at break*=1mm,colback=cellbackground, colframe=cellborder]
\begin{Verbatim}[commandchars=\\\{\}, baselinestretch=1.0]
    {\color{darkgreen}for} M in range(nmax):
       print("  *******   Eigenvalue  #",M{\color{purple}+}{\color{darkgreen}1},"  ******* ")

       f_old{\color{purple}=}{\color{darkgreen}0}
       {\color{darkgreen}for} k in range(kmax): #iteration for eigenvalue candidate
           imatch{\color{purple}=}{\color{darkgreen}0}

{\color{darkgreen}for} j in range(dim{\color{purple}-}1):  #classical right turning point
               xx[{\color{darkgreen}0}]{\color{purple}=}xl;          xx[dim{\color{purple}-}{\color{darkgreen}1}]{\color{purple}=}xu
               DE1 {\color{purple}=} E {\color{purple}-} V(xx[j])
               xx[j{\color{purple}+}{\color{darkgreen}1}] {\color{purple}=} xx[j]{\color{purple}+}h
               DE2 {\color{purple}=} E {\color{purple}-} V(xx[j{\color{purple}+}{\color{darkgreen}1}])
               D1D2{\color{purple}=}DE1{\color{purple}*}DE2
               if D1D2<{\color{purple}=}{\color{darkgreen}0}  and DE1 > {\color{darkgreen}0}:  #match point
                  imatch {\color{purple}=} j{\color{purple}+}{\color{darkgreen}1}
                  print(" imatch {\color{purple}=} ",imatch," xmatch {\color{purple}=} %2.3f"
                        %(xx[imatch])," V(xmatch) {\color{purple}=} %2.5f"
                        %(V(xx[imatch]))," E {\color{purple}=} %2.4f" %(E))
\end{Verbatim}
\end{tcolorbox}
\vspace*{0.1cm}

\vspace*{0.1cm}
\begin{tcolorbox}[breakable, size=fbox, boxrule=1pt, pad at break*=1mm,colback=cellbackground, colframe=cellborder]
\begin{Verbatim}[commandchars=\\\{\}, baselinestretch=1.0]
 xmatch{\color{purple}=}xx[imatch]
           ii{\color{purple}=}range(imatch{\color{purple}+}{\color{darkgreen}2})
           i_lim {\color{purple}=} ii[{\color{darkgreen}2}:imatch{\color{purple}+}{\color{darkgreen}2}]

           xx[{\color{darkgreen}0}]{\color{purple}=}xl;  xx[{\color{darkgreen}1}]{\color{purple}=}xl{\color{purple}+}h   #valores iniciais
           yy[{\color{darkgreen}0}]{\color{purple}=}{\color{darkgreen}0};   yy[{\color{darkgreen}1}] {\color{purple}=} delta

           {\color{darkgreen}for} i in i_lim:      #numerov left solution
               yy[i]{\color{purple}=}nrovl(yy[i{\color{purple}-}2],yy[i{\color{purple}-}1],xx[i{\color{purple}-}2],E,h,iflag)
               xx[i]{\color{purple}=} xx[i{\color{purple}-}1]{\color{purple}+}h

           jjj{\color{purple}=}list(range(dim{\color{purple}+}1))
           j_lim {\color{purple}=} list(jjj[imatch{\color{purple}-}1:dim{\color{purple}+}1])
           comp_j{\color{purple}=}len(j_lim)
           jj{\color{purple}=}sorted(j_lim,key{\color{purple}=}abs,reverse{\color{purple}=}True)

           {\color{darkgreen}for} i in range(dim):
               if i<{\color{purple}=}imatch{\color{purple}+}1:
                  yl[i]{\color{purple}=}yy[i]
               if i>imatch{\color{purple}+}1:
                  yl[i]{\color{purple}=}0

           {\color{darkgreen}for} i in jj:         #numerov right solution
               if i{\color{purple}=}{\color{purple}=}dim:
                   yr[dim{\color{purple}-}1]{\color{purple}=}0
               if i{\color{purple}=}{\color{purple}=}(dim{\color{purple}-}2):
                   yr[dim{\color{purple}-}2]{\color{purple}=}2{\color{purple}*}delta
               if i<(dim{\color{purple}-}2):
                   yr[i]{\color{purple}=}nrovr(yr[i{\color{purple}+}2],yr[i{\color{purple}+}1],xx[i{\color{purple}+}2],E,h,iflag)
                   xx[i]{\color{purple}=} xx[i{\color{purple}+}1]{\color{purple}-}h

           {\color{darkgreen}for} i in range(imatch{\color{purple}-}1):
               yr[i]{\color{purple}=}0

           ymatch{\color{purple}=}yy[imatch]
           yrmatch{\color{purple}=}yr[imatch]
           ylmatch{\color{purple}=}yl[imatch]

           if ymatch !{\color{purple}=} 0:
               scale{\color{purple}=}yrmatch{\color{purple} /}ymatch
           else:
               continue

           {\color{darkgreen}for} t in range(imatch{\color{purple}+}1):                #  y_left
               yy[t] {\color{purple}=} yy[t]{\color{purple}*}scale
               yl[t] {\color{purple}=} {\color{purple}-}yl[t]{\color{purple}*}scale
           yl[imatch{\color{purple}+}1]{\color{purple}=}{\color{purple}-}yl[imatch{\color{purple}+}1]{\color{purple}*}scale
           ymatch{\color{purple}=}yy[imatch]
           dlmatch{\color{purple}=}yy[imatch{\color{purple}+}1]{\color{purple}*}scale{\color{purple}-}yy[imatch{\color{purple}-}1]  # dif1_left

           t_lim{\color{purple}=}list(range(dim{\color{purple}+}1))
           tt{\color{purple}=}list(t_lim[imatch{\color{purple}+}1:dim])

           drmatch{\color{purple}=}yr[imatch{\color{purple}+}1]{\color{purple}-}yr[imatch{\color{purple}-}1]        # dif1_right

           f {\color{purple}=}(dlmatch{\color{purple}-}drmatch){\color{purple} /}(2{\color{purple}*}h)

           {\color{darkgreen}for} t in tt:                             #  y_right
               yy[t] {\color{purple}=} yr[t]



           delta_E{\color{purple}=}{\color{purple}-}f{\color{purple}*}(E{\color{purple}-}E_old){\color{purple} /}(f{\color{purple}-}f_old)


           if abs(delta_E)<eps:  # determinacao da raiz (energia)
                                 # de f(E) pelo metodo da secante
              ee[M] {\color{purple}=} E
              nk[M] {\color{purple}=} k
              ff[M] {\color{purple}=} f
              k {\color{purple}=} kmax
              break
\end{Verbatim}
\end{tcolorbox}
\vspace*{0.1cm}

So far, the only thing we need to change in our program is the equation that defines our effective potential $V(x)$. The rest of the code will be the same for any equation that is in the same form as equation~(\ref{eq9}). From now on, we must change the code whenever we are looking for solutions with a different potential.

The parameter $M$ indicates the eigenvalue that we are determining, in the next step, in order to avoid that the program needs to sweep the entire potential well in search of solutions, we can give increments between one eigenvalue and another in the form of multiples of the $\mbox{d}E$ parameter, which will be duly defined in a next step.

Usually, this process involves trial and error, where we adjust the $\mbox{d}E$ multiplier for each case, until we find the desired eigenvalue. However, as we are developing this first example for the hydrogen atom, which already has its energy eigenvalues well defined, this task becomes much easier. Before we find the ground state ($M=0$) we must use the step $24 \times \mbox{d}E$. Thus, we find for ground state energy the value of $\varepsilon = -0.99453$, which must be compared with the known value of the ground state energy for the hydrogen atom $\varepsilon = -1$. And the next multipliers so that we can find at least the first three solutions witch are:

\vspace*{0.1cm}
\begin{tcolorbox}[breakable, size=fbox, boxrule=1pt, pad at break*=1mm,colback=cellbackground, colframe=cellborder]
\begin{Verbatim}[commandchars=\\\{\}, baselinestretch=1.0]
           else:
              f_old{\color{purple}=}f;  E_old{\color{purple}=}E;   E {\color{purple}=} E {\color{purple}+} 24{\color{purple}*}dE

       if M{\color{purple}=}{\color{purple}=}0:
       	  E {\color{purple}=} E {\color{purple}+} 126{\color{purple}*}dE
          {\color{darkgreen}for} j in range(dim):
              yy1[j]{\color{purple}=}yy[j]
       if M{\color{purple}=}{\color{purple}=}1:
          E {\color{purple}=} E {\color{purple}+} 3{\color{purple}*}dE
          {\color{darkgreen}for} j in range(dim):
              yy2[j]{\color{purple}=}yy[j]
       if M{\color{purple}=}{\color{purple}=}2:
          E {\color{purple}=} E_old {\color{purple}+} 6{\color{purple}*}dE
          {\color{darkgreen}for} j in range(dim):
              yy3[j]{\color{purple}=}yy[j]
       print(" k {\color{purple}=} %2d   E_old {\color{purple}=} %2.3f    Eingen {\color{purple}=} %2.3f
               f_old {\color{purple}=} %2.3e   f {\color{purple}=} %2.3e   delta_E {\color{purple}=} %2.3e"
               %(nk[M],E_old,ee[M],f_old,f,delta_E))
       print()

    {\color{darkgreen}return} ee, xx, yy1, yy2, yy3
\end{Verbatim}
\end{tcolorbox}
\vspace*{0.1cm}

If you are interested in finding other solutions, you should add the steps for large values of $M$.

We finally reach the part of the program where we must introduce the initial values to proceed with the solution. Whenever we start to develop a new code, these must be the first values to be changed, right after choosing the effective potential $V(x)$.

The parameters $a$ and $b$ correspond to the boundary points mentioned in the second section. At first, our wave function should have its initial value ($a$) equal to 0, however, to avoid divisions by 0 throughout the program we should start from a relatively small value ($a=0.001$). In the piece of code below $h$ represents the $\delta$ increment, $kmax$ is the maximum number of interactions for each eigenvalue, $nmax$ is the number of eigenvalues that we are trying to find, $Ein$ is the minimum energy of de effective potential that we are trying to solve, and $Vmax$ is the maximum energy of the same effective potential.
\vspace*{0.1cm}
\begin{tcolorbox}[breakable, size=fbox, boxrule=1pt, pad at break*=1mm,colback=cellbackground, colframe=cellborder]
\begin{Verbatim}[commandchars=\\\{\}, baselinestretch=1.0]
a{\color{purple}=}0.001; b{\color{purple}=} 60.001; h{\color{purple}=}0.01
xl{\color{purple}=}a; xu{\color{purple}=}b; D {\color{purple}=} xu{\color{purple}-}xl
delta {\color{purple}=} 0.02; eps {\color{purple}=} 0.0001
dim{\color{purple}=}int(D{\color{purple} /}h); kmax{\color{purple}=}300; nmax{\color{purple}=}3
n {\color{purple}=}0; iflag{\color{purple}=}0
Rydberg{\color{purple}=}13.605693122994

x0{\color{purple}=}xl; y0{\color{purple}=}0. ; y1{\color{purple}=}delta; iflag{\color{purple}=}1
nrovl(y0,y1,x0,0,h,iflag)
nrovr(y0,y1,xu,0,h,iflag)

dE {\color{purple}=} delta{\color{purple} /}4
Ein {\color{purple}=} {\color{purple}-}1.6
Vmax{\color{purple}=} 0.
\end{Verbatim}
\end{tcolorbox}
\vspace*{0.1cm}

With the inputs given in the previous step, we can now print the initial values of our program on the screen.

\vspace*{0.1cm}
\begin{tcolorbox}[breakable, size=fbox, boxrule=1pt, pad at break*=1mm,colback=cellbackground, colframe=cellborder]
\begin{Verbatim}[commandchars=\\\{\}, baselinestretch=1.0]
print(" =================================================")
print("  ")
print("       Potential : V(x) {\color{purple}=} {\color{purple}-}2{\color{purple} /}x   (Hydrogen L=0)   ")
print("  ")
print(" =================================================")

print(" E_in {\color{purple}=} %2.4f" %(Ein)," dE {\color{purple}=} ",dE," h {\color{purple}=} ",h," dim {\color{purple}=} ",dim)
print("  ")

corlors {\color{purple}=}['b','r','g','m','c']

ee,xx,yy1,yy2,yy3{\color{purple}=}espectro(xl,xu,h,delta,eps,dim,nmax,kmax,
                                           Ein,Vmax,dE,iflag)

A{\color{purple}=}1.                     # amplitude normalization in 1 unit
ymax1{\color{purple}=}vec_max(dim,yy1)
ymax2{\color{purple}=}vec_max(dim,yy2)
ymax3{\color{purple}=}vec_max(dim,yy3)

{\color{darkgreen}for} i in range(dim):     # amplitude normalization
    yy1[i] {\color{purple}=} yy1[i]{\color{purple} /}ymax1;  yy2[i] {\color{purple}=} yy2[i]{\color{purple} /}ymax2
    yy3[i] {\color{purple}=} yy3[i]{\color{purple} /}ymax3

\end{Verbatim}
\end{tcolorbox}
\vspace*{0.1cm}

\begin{figure}[ht]
\centerline{\includegraphics[width=12.0cm]{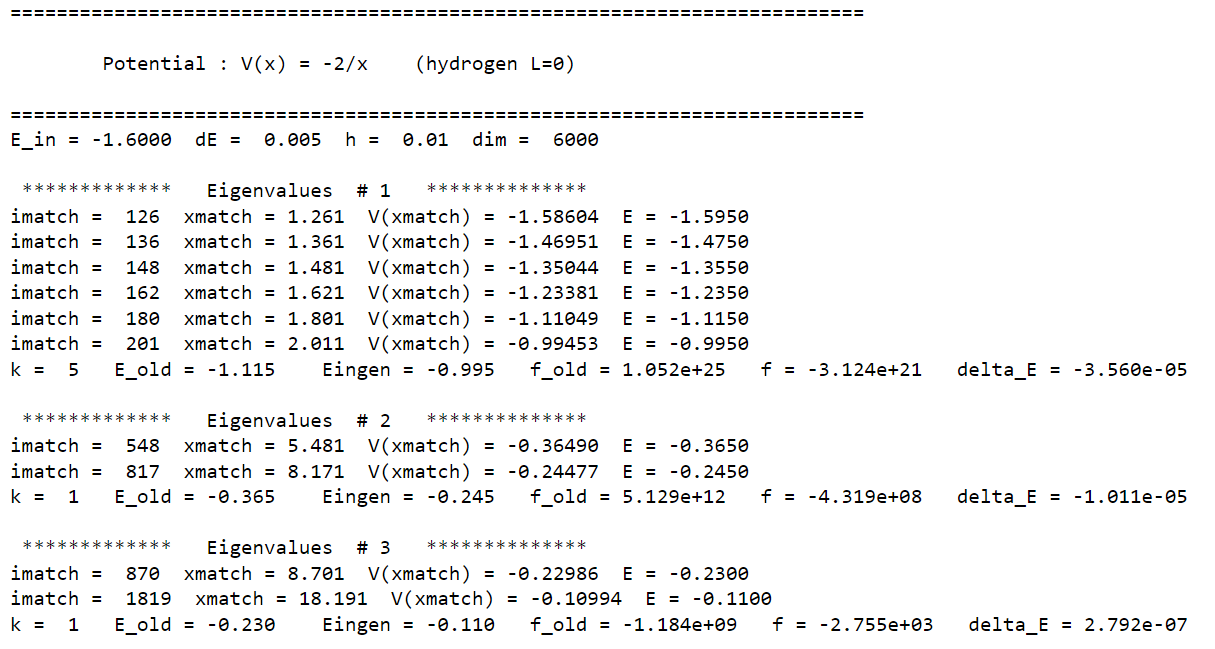}}
\caption{Output of eigenvalues displayed by the code.}
\label{fig2}
\end{figure}

Finally, at this point the program was able to determine the eigenvalues and eigenfunctions associated with the first three states of the hydrogen atom. Initially, the program prints the eigenvalues ($Eingen$) on the screen according to the Figure~\ref{fig2}.

\newpage
Table~\ref{tabela1} shows the comparison between our results, extract from the Figure~\ref{fig2}, and the well known analytical values for the hydrogen atom, in Rydberg units, given by the formula $E_n = -1/n^2$.

\renewcommand{\arraystretch}{0.9}
\begin{table}[ht]
  \caption{Comparison between the energy values, in Rydberg units, found by the Numerov's method and the analytical ones for $\ell=0$ and n=1, 2 and 3.}\label{tabela1}
  \vspace*{0.2cm}
  \begin{center}
  \begin{tabular}{c|c|c}
    \hline
    \hline
      n           &Numerov's Energy    &Analytical value   \\ \hline
      1           & -0.995             & -1    \\
      2           &-0.245              & -0.25   \\
      3           &-0.110              & -0.111  \\

      \hline
      \hline
  \end{tabular}
  \end{center}
\end{table}
\renewcommand{\arraystretch}{1}

From now on we will introduce the codes necessary to generate the graphs, as well as calculate the normalization of the wavefunctions. First, let's plot the effective potential graph. During the process of setting the code for different potentials, it is important to know the effective potential in order to adjust the parameters accordingly.

\vspace*{0.1cm}
\begin{tcolorbox}[breakable, size=fbox, boxrule=1pt, pad at break*=1mm,colback=cellbackground, colframe=cellborder]
\begin{Verbatim}[commandchars=\\\{\}, baselinestretch=1.0]

print(" ")
print(" %60s" %('Potential'))
figure(figsize{\color{purple}=}(8, 6), dpi{\color{purple}=}800)
plot(x,V(x),'k{\color{purple}-}',linewidth{\color{purple}=}2)
ylim({\color{purple}-}10,1)
xlim(a,10)
xlabel('x')
ylabel('Effective potential')
grid()
show())
\end{Verbatim}
\end{tcolorbox}
\vspace*{0.1cm}

\begin{figure}[ht]
\centerline{\includegraphics[width=12.0cm]{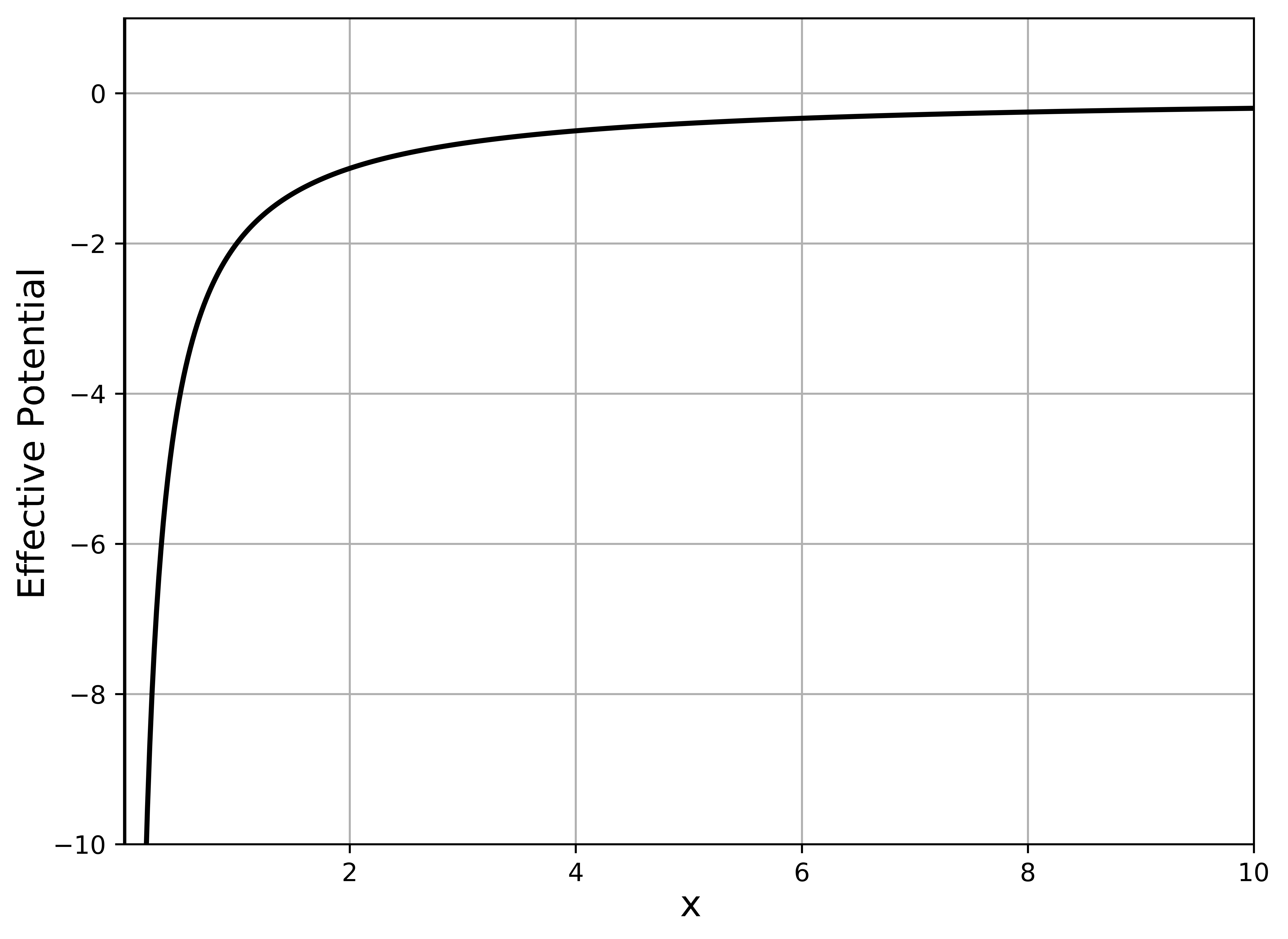}}
\caption{effective potential of equation~(\ref{sch2}) for $\ell=0$.}
\label{fig3}
\end{figure}

\noindent The code above plots the effective potential, as we can see in figure~\ref{fig3}

And the graph for the eigenfunctions with an arbitrary normalization, is built by

\vspace*{0.1cm}
\begin{tcolorbox}[breakable, size=fbox, boxrule=1pt, pad at break*=1mm,colback=cellbackground, colframe=cellborder]
\begin{Verbatim}[commandchars=\\\{\}, baselinestretch=1.0]
print(" ")
print(" %60s" %(''))
figure(figsize{\color{purple}=}(8, 6), dpi{\color{purple}=}800)
plot(xx,yy1,'b{\color{purple}-}',linewidth{\color{purple}=}1)
plot(xx,yy2,'r{\color{purple}-}')
plot(xx,yy3,'g{\color{purple}-}')
legend([' y1','  y2','  y3'],prop{\color{purple}=}{"size":10},frameon{\color{purple}=}False)
ylim({\color{purple}-}0.7,1.1)
xlim(0,40)
xlabel('x')
ylabel('Eigenfunctions')
grid()
show()
\end{Verbatim}
\end{tcolorbox}
\vspace*{0.1cm}

\begin{figure}[ht]
\centerline{\includegraphics[width=12.0cm]{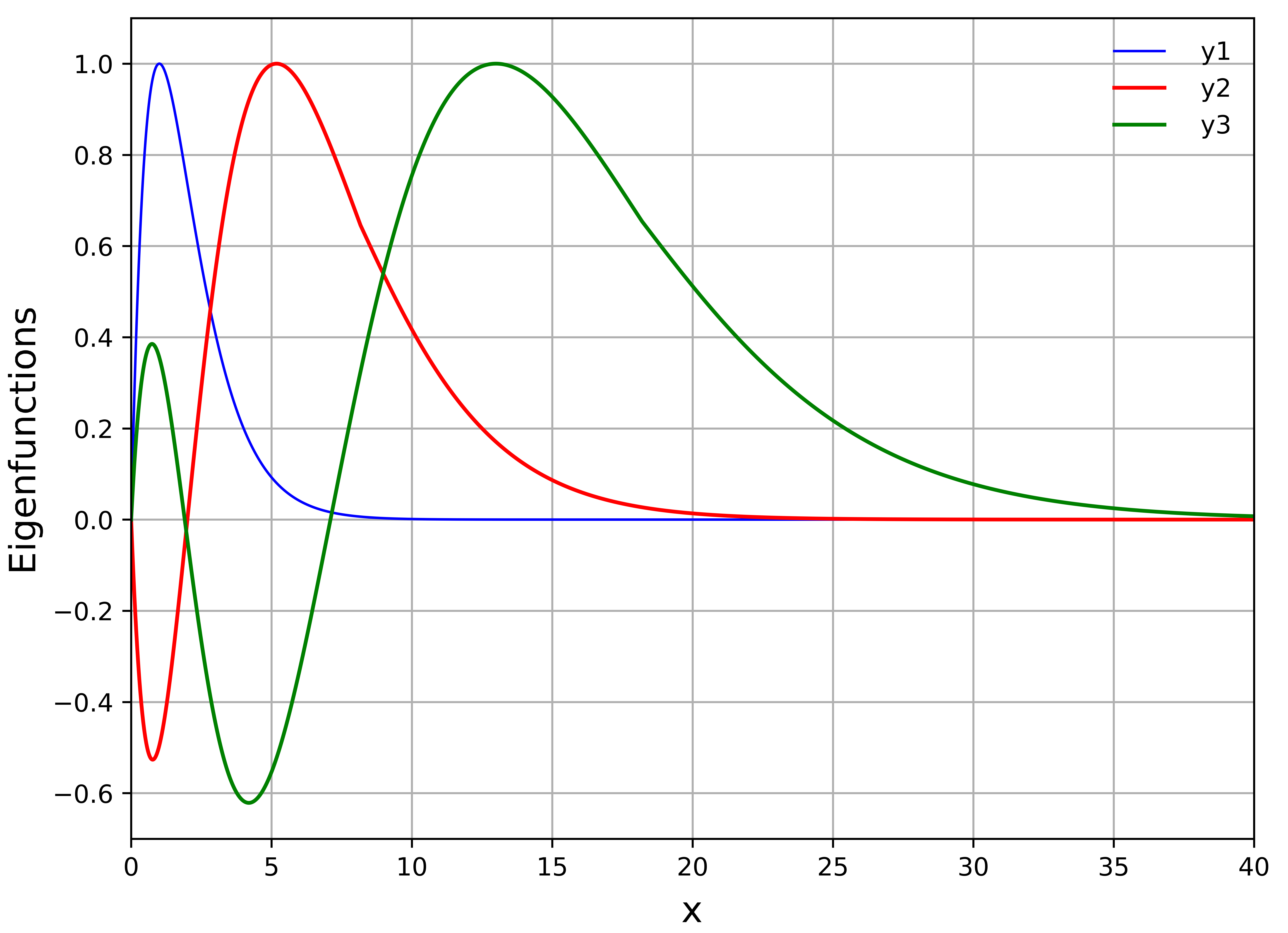}}
\caption{First three eigenfunctions of equation~(\ref{sch2}), with $\ell=0$, calculated with the Numerov method.}
\label{fig4}
\end{figure}

\noindent Which generates the plot shown in the figure~\ref{fig4}

As a last step, let's include in our code, the calculation of the normalization of the wave functions, given by:

\vspace*{0.1cm}
\begin{tcolorbox}[breakable, size=fbox, boxrule=1pt, pad at break*=1mm,colback=cellbackground, colframe=cellborder]
\begin{Verbatim}[commandchars=\\\{\}, baselinestretch=1.0]
efic{\color{purple}=}list(range(nmax));  I{\color{purple}=}list(range(nmax))
eI{\color{purple}=}list(range(nmax))

S{\color{purple}=}A{\color{purple}*}(xu{\color{purple}-}xl)                     # Born normalization
l0{\color{purple}=}0; l1{\color{purple}=}0; l2{\color{purple}=}0
N{\color{purple}=}10000
{\color{darkgreen}for} i in range(N):              # integral de y{\color{purple}*}y
    y {\color{purple}=} A{\color{purple}*}random(1)
    j{\color{purple}=}randint(dim)
    yj1{\color{purple}=}yy1[j]{\color{purple}*}yy1[j];   yj2{\color{purple}=}yy2[j]{\color{purple}*}yy2[j];   yj3{\color{purple}=}yy3[j]{\color{purple}*}yy3[j]
    if y <{\color{purple}=} yj1:
       l0 {\color{purple}+}{\color{purple}=} 1
    if y <{\color{purple}=} yj2:
       l1 {\color{purple}+}{\color{purple}=} 1
    if y <{\color{purple}=} yj3:
       l2 {\color{purple}+}{\color{purple}=} 1

efic[0] {\color{purple}=} float(l0){\color{purple} /}N;  efic[1] {\color{purple}=} float(l1){\color{purple} /}N
efic[2] {\color{purple}=} float(l2){\color{purple} /}N
I[0] {\color{purple}=} S{\color{purple}*}efic[0];       I[1] {\color{purple}=} S{\color{purple}*}efic[1];       I[2] {\color{purple}=} S{\color{purple}*}efic[2]
eI[0] {\color{purple}=} (S{\color{purple} /}sqrt(N)){\color{purple}*}sqrt(efic[0]{\color{purple}*}(1{\color{purple}-}efic[0]))
eI[1] {\color{purple}=} (S{\color{purple} /}sqrt(N)){\color{purple}*}sqrt(efic[1]{\color{purple}*}(1{\color{purple}-}efic[1]))
eI[2] {\color{purple}=} (S{\color{purple} /}sqrt(N)){\color{purple}*}sqrt(efic[2]{\color{purple}*}(1{\color{purple}-}efic[2]))

print("  ")
print("  probability normalization ")
print("efic1{\color{purple}=}%2.3f I1{\color{purple}=}%2.3f eI1 {\color{purple}=} %2.3f" %(efic[0],I[0],eI[0]))
print("efic2{\color{purple}=}%2.3f I2{\color{purple}=}%2.3f eI2 {\color{purple}=} %2.3f" %(efic[1],I[1],eI[1]))
print("efic2{\color{purple}=}%2.3f I2{\color{purple}=}%2.3f eI2 {\color{purple}=} %2.3f" %(efic[2],I[2],eI[2]))
\end{Verbatim}
\end{tcolorbox}
\vspace*{0.1cm}

\vspace*{0.1cm}
\begin{tcolorbox}[breakable, size=fbox, boxrule=1pt, pad at break*=1mm,colback=cellbackground, colframe=cellborder]
\begin{Verbatim}[commandchars=\\\{\}, baselinestretch=1.0]
{\color{darkgreen}for} i in range(dim): # probability normalization
    yy1[i]{\color{purple}=}yy1[i]{\color{purple} /}sqrt(I[0]); yy2[i]{\color{purple}=}yy2[i]{\color{purple} /}sqrt(I[1])
    yy3[i]{\color{purple}=}yy3[i]{\color{purple} /}sqrt(I[2])


ymax1{\color{purple}=}vec_max(dim,yy1);   ymax2{\color{purple}=}vec_max(dim,yy2)
ymax3{\color{purple}=}vec_max(dim,yy3)

S0{\color{purple}=}ymax1{\color{purple}*}ymax1{\color{purple}*}(xu{\color{purple}-}xl);   S1{\color{purple}=}ymax2{\color{purple}*}ymax2{\color{purple}*}(xu{\color{purple}-}xl)
S2{\color{purple}=}ymax3{\color{purple}*}ymax3{\color{purple}*}(xu{\color{purple}-}xl)

l0{\color{purple}=}0;  l1{\color{purple}=}0; l2{\color{purple}=}0           # checking probability
N{\color{purple}=}10000
{\color{darkgreen}for} i in range(N):
    y1 {\color{purple}=} ymax1{\color{purple}*}ymax1{\color{purple}*}random(1); y2 {\color{purple}=} ymax2{\color{purple}*}ymax2{\color{purple}*}random(1)
    y3 {\color{purple}=} ymax3{\color{purple}*}ymax3{\color{purple}*}random(1)
    j{\color{purple}=}randint(dim)
    yj1{\color{purple}=}yy1[j]{\color{purple}*}yy1[j];   yj2{\color{purple}=}yy2[j]{\color{purple}*}yy2[j];    yj3{\color{purple}=}yy3[j]{\color{purple}*}yy3[j]
    if y1<{\color{purple}=} yj1:
       l0 {\color{purple}+}{\color{purple}=} 1
    if y2 <{\color{purple}=} yj2:
       l1 {\color{purple}+}{\color{purple}=} 1
    if y3 <{\color{purple}=} yj3:
       l2 {\color{purple}+}{\color{purple}=} 1

efic[0] {\color{purple}=} float(l0){\color{purple} /}N;  efic[1] {\color{purple}=} float(l1){\color{purple} /}N
efic[2] {\color{purple}=} float(l2){\color{purple} /}N
I[0] {\color{purple}=} S0{\color{purple}*}efic[0];    I[1] {\color{purple}=} S1{\color{purple}*}efic[1];    I[2] {\color{purple}=} S2{\color{purple}*}efic[2]
eI[0] {\color{purple}=} (S0{\color{purple} /}sqrt(N)){\color{purple}*}sqrt(efic[0]{\color{purple}*}(1{\color{purple}-}efic[0]))
eI[1] {\color{purple}=} (S1{\color{purple} /}sqrt(N)){\color{purple}*}sqrt(efic[1]{\color{purple}*}(1{\color{purple}-}efic[1]))
eI[2] {\color{purple}=} (S2{\color{purple} /}sqrt(N)){\color{purple}*}sqrt(efic[2]{\color{purple}*}(1{\color{purple}-}efic[2]))

print("  ")
print("  checking probability ")
print(" efic1{\color{purple}=}%2.3f  I1{\color{purple}=}%2.3f   eI1{\color{purple}=}%2.3f" %(efic[0],I[0],eI[0]))
print(" efic2{\color{purple}=}%2.3f  I2{\color{purple}=}%2.3f   eI2{\color{purple}=}%2.3f" %(efic[1],I[1],eI[1]))
print(" efic3{\color{purple}=}%2.3f  I3{\color{purple}=}%2.3f   eI3{\color{purple}=}%2.3f" %(efic[2],I[2],eI[2]))

\end{Verbatim}
\end{tcolorbox}
\vspace*{0.1cm}

\vspace*{0.1cm}
\begin{tcolorbox}[breakable, size=fbox, boxrule=1pt, pad at break*=1mm,colback=cellbackground, colframe=cellborder]
\begin{Verbatim}[commandchars=\\\{\}, baselinestretch=1.0]

print(" %70s " %(''))

colors {\color{purple}=}['b','r','g','m','c']
figure(figsize{\color{purple}=}(8, 6), dpi{\color{purple}=}800)
plot(xx,yy1,color{\color{purple}=}cores[0])
plot(xx,yy2,color{\color{purple}=}cores[1])
plot(xx,yy3,color{\color{purple}=}cores[2])
#plot(xx,yy4,color{\color{purple}=}cores[3])
#plot(xx,yy5,color{\color{purple}=}cores[4])
legend(['y1','y2','y3'],prop{\color{purple}=}{"size":10},frameon{\color{purple}=}False)
ylim({\color{purple}-}0.3,0.8)
xlim(0,40)
xlabel('x')
ylabel('Normalized Eigenfunctions')
grid(True)
show()

\end{Verbatim}
\end{tcolorbox}
\vspace*{0.1cm}

Thus, we were able to obtain our final graph, composed of the first three wave functions for the hydrogen atom.

\begin{figure}[ht]
\centerline{\includegraphics[width=12.0cm]{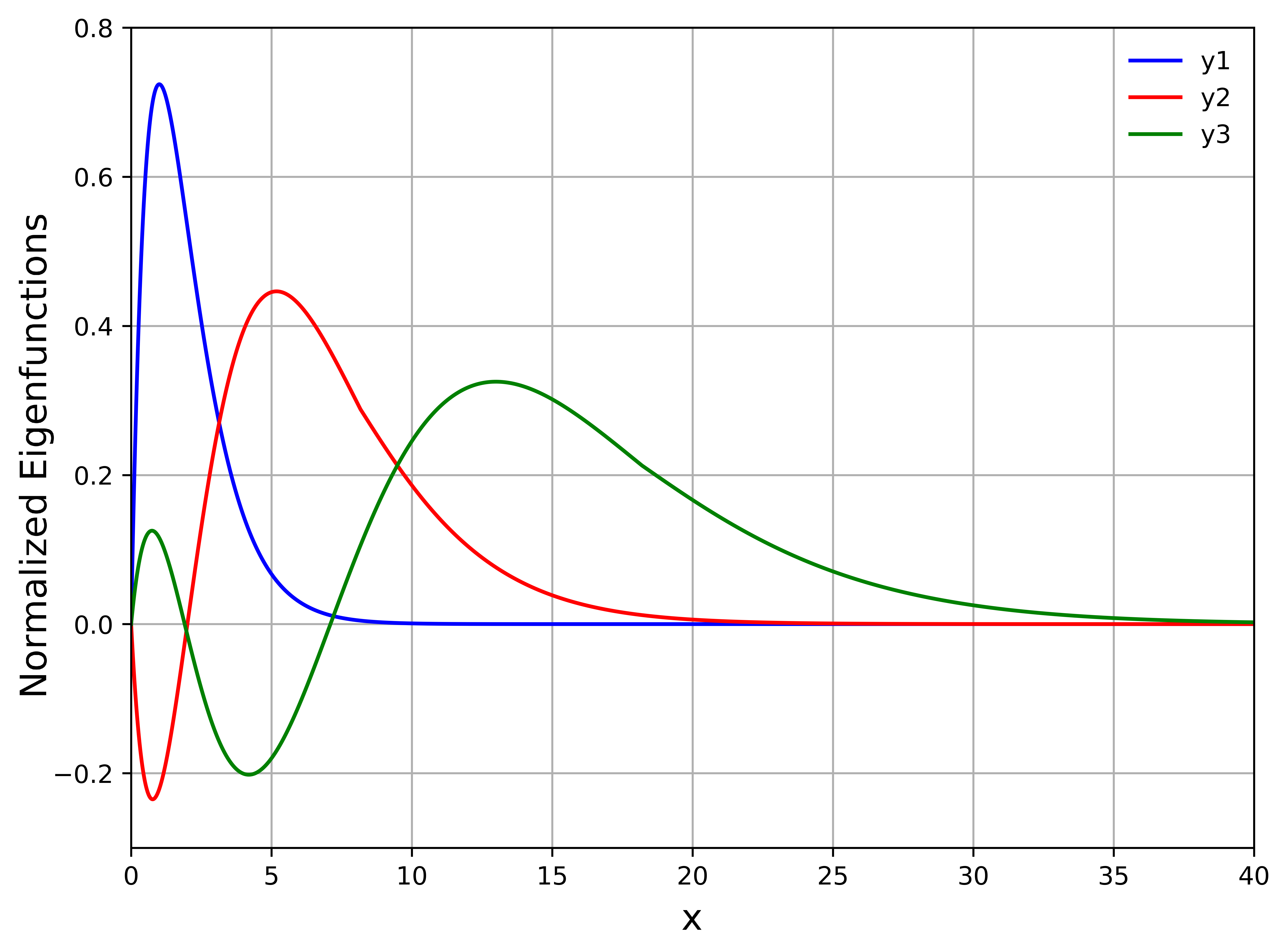}}
\caption{First three normalized eigenfunctions of equation~(\ref{sch2}), with $\ell=0$, calculated with the Numerov method.}
\label{fig42}
\end{figure}

\newpage
\section{Morse Potential}

The Morse potential is a common model for the interatomic interaction of a diatomic molecule~\cite{morse, morse2}. In this section, in order to learn how we can use the Phyton code in other problems, we will see what we must change in the code that was made available in the previous section so that the program will be able to solve equation~(\ref{sch1}) for the quantum number $\ell=0$ and the Morse potential with arbitrary parameters given by:

\begin{equation}\label{morse}
  V(x) = 16 \left(1-e^{-2x} \right)^2
\end{equation}

In this example, we will calculate the first two bounded states, for that, we must adjust the $\mbox{d}E$ multiplier for each case as:

\vspace*{0.1cm}
\begin{tcolorbox}[breakable, size=fbox, boxrule=1pt, pad at break*=1mm,colback=cellbackground, colframe=cellborder]
\begin{Verbatim}[commandchars=\\\{\}, baselinestretch=1.0]
           else:
              f_old{\color{purple}=}f;  E_old{\color{purple}=}E;   E {\color{purple}=} E {\color{purple}+} 7.2{\color{purple}*}dE

       if m{\color{purple}=}{\color{purple}=}0:
       	  E {\color{purple}=} E {\color{purple}+} 70{\color{purple}*}dE
          {\color{darkgreen}for} j in range(dim):
              yy1[j]{\color{purple}=}yy[j]
       if m{\color{purple}=}{\color{purple}=}1:
          E {\color{purple}=} E {\color{purple}+} dE{\color{purple}/}10
          {\color{darkgreen}for} j in range(dim):
              yy2[j]{\color{purple}=}yy[j]
       if m{\color{purple}>}{\color{purple}=}2:
          E {\color{purple}=} E_old {\color{purple}+} 50{\color{purple}*}dE
          {\color{darkgreen}for} j in range(dim):
              yy3[j]{\color{purple}=}yy[j]
       print(" k {\color{purple}=} %2d   E_old {\color{purple}=} %2.3f    Eingen {\color{purple}=} %2.3f
               f_old {\color{purple}=} %2.3e   f {\color{purple}=} %2.3e   delta_E {\color{purple}=} %2.3e"
               %(nk[m],E_old,ee[m],f_old,f,delta_E))
       print()

    {\color{darkgreen}return} ee, xx, yy1, yy2, yy3
\end{Verbatim}
\end{tcolorbox}
\vspace*{0.1cm}

And, the most important part, which is the adjustment of the initial data of our problem. Analyzing the effective potential we can verify that the value os parameter $a$ must be negative, and its minimum value is zero, so the code must be set as follows:

\vspace*{0.1cm}
\begin{tcolorbox}[breakable, size=fbox, boxrule=1pt, pad at break*=1mm,colback=cellbackground, colframe=cellborder]
\begin{Verbatim}[commandchars=\\\{\}, baselinestretch=1.0]
a{\color{purple}=}-1.01; b{\color{purple}=} 5.01; h{\color{purple}=}0.006
xl{\color{purple}=}a; xu{\color{purple}=}b; D {\color{purple}=} xu{\color{purple}-}xl
delta {\color{purple}=} 0.01; eps {\color{purple}=} 0.00001
dim{\color{purple}=}int(D{\color{purple} /}h); kmax{\color{purple}=}100; nmax{\color{purple}=}2
n {\color{purple}=}0; iflag{\color{purple}=}0
Rydberg{\color{purple}=}13.605693122994

x0{\color{purple}=}xl; y0{\color{purple}=}0. ; y1{\color{purple}=}delta; iflag{\color{purple}=}1
nrovl(y0,y1,x0,0,h,iflag)
nrovr(y0,y1,xu,0,h,iflag)

dE {\color{purple}=} delta
Ein {\color{purple}=} 0.0
Vmax{\color{purple}=} 16.
\end{Verbatim}
\end{tcolorbox}
\vspace*{0.1cm}

From now on, the program is able to find the energies of the first two states, which are respectively $7.1380$ and $15.0380$. As well as already determined the wave functions also for the first two states.

\begin{figure}[htb]
\centerline{\includegraphics[width=12.0cm]{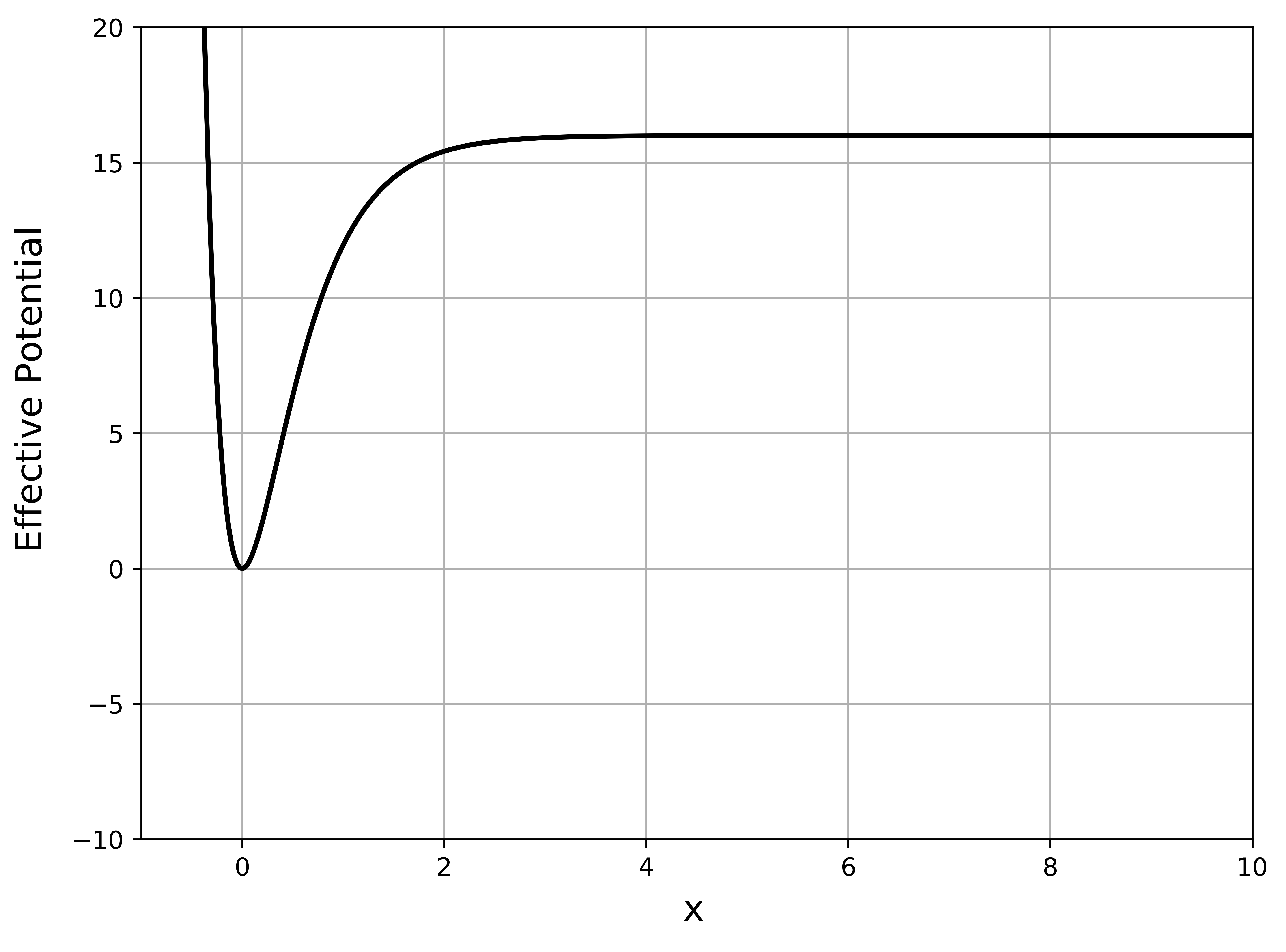}}
\caption{Effective Potential given by equation~(\ref{morse}).}
\label{fig5}
\end{figure}

\begin{figure}[htb]
\centerline{\includegraphics[width=12.0cm]{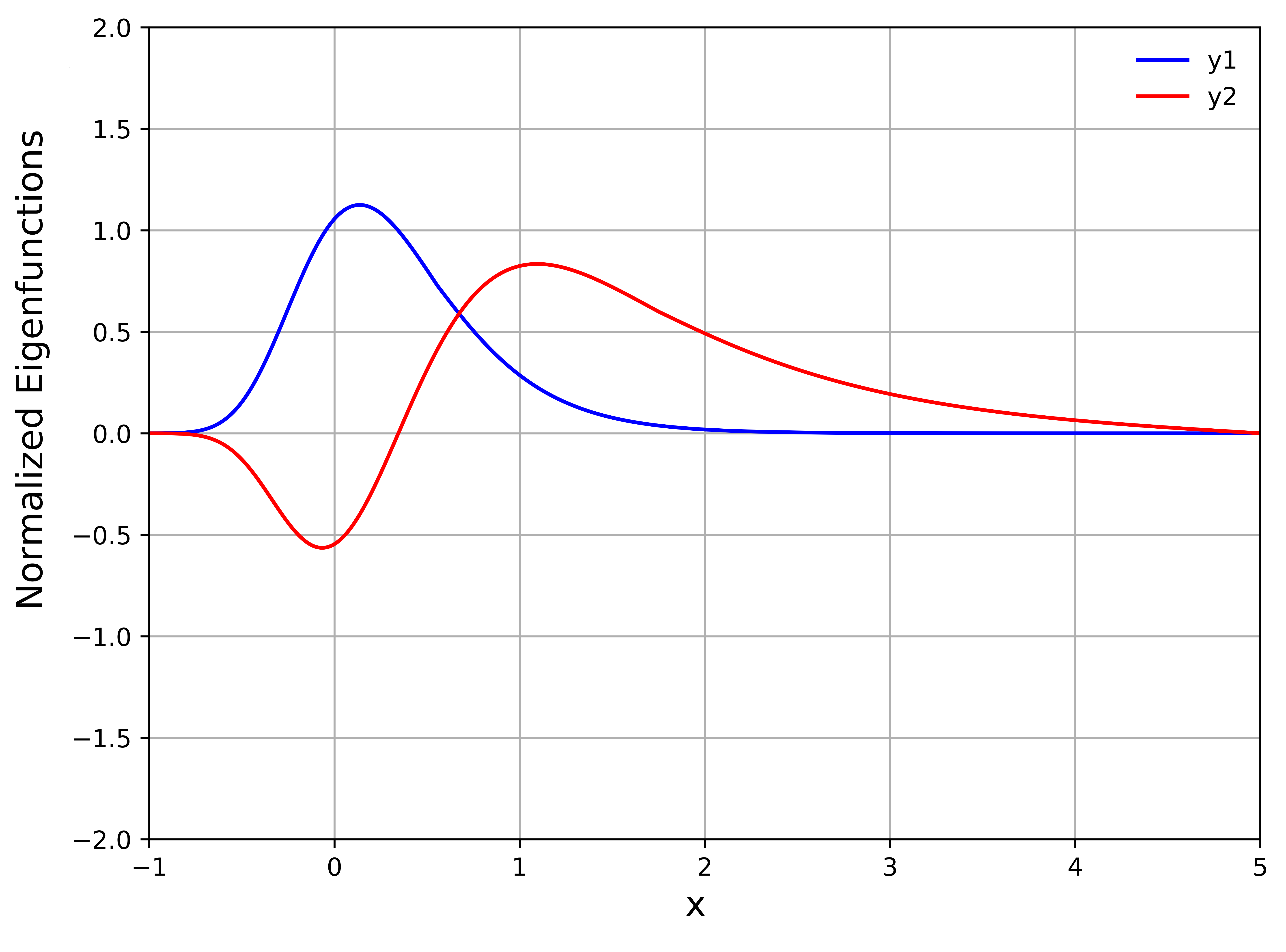}}
\caption{Normalized Eigenfunctions of equation~(\ref{sch2}), with the potential given by equation~(\ref{morse}) with $\ell=0$.}
\label{fig6}
\end{figure}

Then, all that remains is to get all the aesthetic part of the code right, adjusting the limits of graphics, subtitles and other factors. Thus we first get the effective potential, as shown in Figure~\ref{fig5} and finally arrive at the normalized wave functions shown in Figure~\ref{fig6}.
\newpage
In Table~\ref{tabela2}, we can compare the eigenvalues found by the Numerov's method with the analytical values given by $E_n = 16 \left[\left(n+\frac{1}{2}\right)-\frac{1}{4}\left(n+\frac{1}{2}\right)\right]$.

\renewcommand{\arraystretch}{0.9}
\begin{table}[ht]
  \caption{Comparison between the energy values for the Morse potential, in Rydberg units, found by the Numerov's method and the analytical ones for $\ell=0$ and n=0 and 1.}\label{tabela2}
  \vspace*{0.2cm}
  \begin{center}
  \begin{tabular}{c|c|c}
    \hline
    \hline
      n           &Numerov's Energy    &Analytical value   \\ \hline
      0           & 7.1380             & 7.0  \\
      1           & 15.0380              & 15.0   \\
      \hline
      \hline
  \end{tabular}
  \end{center}
\end{table}
\renewcommand{\arraystretch}{1}

\newpage
\section{Quantum Dot}

The development of technology based on quantum dots is quite recent, but it is already showing signs that it is the next great technology, when we talk about optics. In a simple model for a quantum dot
composed of two electrons, they can be described with a external harmonic oscillator potential of frequency $\Omega = 2\omega$. Following the steps of reference~\cite{BJP}, we have the
effective potential to be introduced into equation~(\ref{sch1}) for the quantum number $\ell=0$ is given by:

\begin{equation}\label{qd}
  V(x) = \frac{1}{x} + \omega^2 x - \frac{0.25}{x^2}
\end{equation}

\noindent introducing this potential into our code, let's now calculate the first five solutions that the program is capable of finding. We can observe that in our code used so far, we only have up to three wave functions,
in this example we will show how we included two more solutions in the code. The procedure is very simple, just add the $yy4$ and $yy5$ functions along the code and all the other parts that are related to them, for example,

\vspace*{0.1cm}
\begin{tcolorbox}[breakable, size=fbox, boxrule=1pt, pad at break*=1mm,colback=cellbackground, colframe=cellborder]
\begin{Verbatim}[commandchars=\\\{\}, baselinestretch=1.0]
    xx{\color{purple}=}list(range(dim));     yy{\color{purple}=}list(range(dim))
    ww{\color{purple}=}list(range(dim));     yl{\color{purple}=}list(range(dim))
    yr{\color{purple}=}list(range(dim));     ee{\color{purple}=}list(range(nmax))
    ff{\color{purple}=}list(range(nmax));    ff2{\color{purple}=}list(range(nmax))
    yy1{\color{purple}=}list(range(dim));    yy2{\color{purple}=}list(range(dim))
    yy3{\color{purple}=}list(range(dim));   {\color{red} yy4{\color{purple}=}list(range(dim))}
    {\color{red} yy5{\color{purple}=}list(range(dim))}
    colors {\color{purple}=}['b','r','g','m','c'];  nk{\color{purple}=}list(range(nmax))
    E_old {\color{purple}=} Ein;   E {\color{purple}=} Ein {\color{purple}+} dE
\end{Verbatim}
\end{tcolorbox}
\vspace*{0.1cm}

Now, as in the previous example, let's include the $dE$ multipliers, remembering to include the new wave functions.

\vspace*{0.1cm}
\begin{tcolorbox}[breakable, size=fbox, boxrule=1pt, pad at break*=1mm,colback=cellbackground, colframe=cellborder]
\begin{Verbatim}[commandchars=\\\{\}, baselinestretch=1.0]
           else:
              f_old{\color{purple}=}f;  E_old{\color{purple}=}E;   E {\color{purple}=} E {\color{purple}+} dE{\color{purple}/}25

       if m{\color{purple}=}{\color{purple}=}0:
          E {\color{purple}=} E {\color{purple}+} 6{\color{purple}*}dE
          {\color{darkgreen}for} j in range(dim):
              yy1[j]{\color{purple}=}yy[j]
       if m{\color{purple}=}{\color{purple}=}1:
          E {\color{purple}=} E {\color{purple}+} 6{\color{purple}*}dE
          {\color{darkgreen}for} j in range(dim):
              yy2[j]{\color{purple}=}yy[j]
       if m{\color{purple}>}{\color{purple}=}2:
          E {\color{purple}=} E_old {\color{purple}+} 13{\color{purple}*}dE{\color{purple}/}2
          {\color{darkgreen}for} j in range(dim):
              yy3[j]{\color{purple}=}yy[j]
       if m{\color{purple}>}{\color{purple}=}3:
          E {\color{purple}=} E_old {\color{purple}+} 13{\color{purple}*}dE{\color{purple}/}2
          {\color{darkgreen}for} j in range(dim):
              yy4[j]{\color{purple}=}yy[j]
       if m{\color{purple}>}{\color{purple}=}4:
          E {\color{purple}=} E_old {\color{purple}+} 4{\color{purple}*}dE
          {\color{darkgreen}for} j in range(dim):
              yy5[j]{\color{purple}=}yy[j]
       print(" k {\color{purple}=} %2d   E_old {\color{purple}=} %2.3f    Eingen {\color{purple}=} %2.3f
               f_old {\color{purple}=} %2.3e   f {\color{purple}=} %2.3e   delta_E {\color{purple}=} %2.3e"
               %(nk[m],E_old,ee[m],f_old,f,delta_E))
       print()

    {\color{darkgreen}return} ee, xx, yy1, yy2, yy3. yy4, yy5
\end{Verbatim}
\end{tcolorbox}
\vspace*{0.1cm}

And finally, we must include the initial conditions for our problem, as shown below.

\vspace*{0.1cm}
\begin{tcolorbox}[breakable, size=fbox, boxrule=1pt, pad at break*=1mm,colback=cellbackground, colframe=cellborder]
\begin{Verbatim}[commandchars=\\\{\}, baselinestretch=1.0]
a{\color{purple}=}0.001; b{\color{purple}=} 60.001; h{\color{purple}=}0.02
xl{\color{purple}=}a; xu{\color{purple}=}b; D {\color{purple}=} xu{\color{purple}-}xl
delta {\color{purple}=} 0.01; eps {\color{purple}=} 0.00001
dim{\color{purple}=}int(D{\color{purple} /}h); kmax{\color{purple}=}100; nmax{\color{purple}=}5
n {\color{purple}=}0; iflag{\color{purple}=}0
Rydberg{\color{purple}=}13.605693122994

x0{\color{purple}=}xl; y0{\color{purple}=}0. ; y1{\color{purple}=}delta; iflag{\color{purple}=}1
nrovl(y0,y1,x0,0,h,iflag)
nrovr(y0,y1,xu,0,h,iflag)

dE {\color{purple}=} delta{\color{purple}/}2.8
Ein {\color{purple}=} 0.086857
Vmax{\color{purple}=} 2.
\end{Verbatim}
\end{tcolorbox}
\vspace*{0.1cm}

\begin{figure}[hbt]
\centerline{\includegraphics[width=9.0cm]{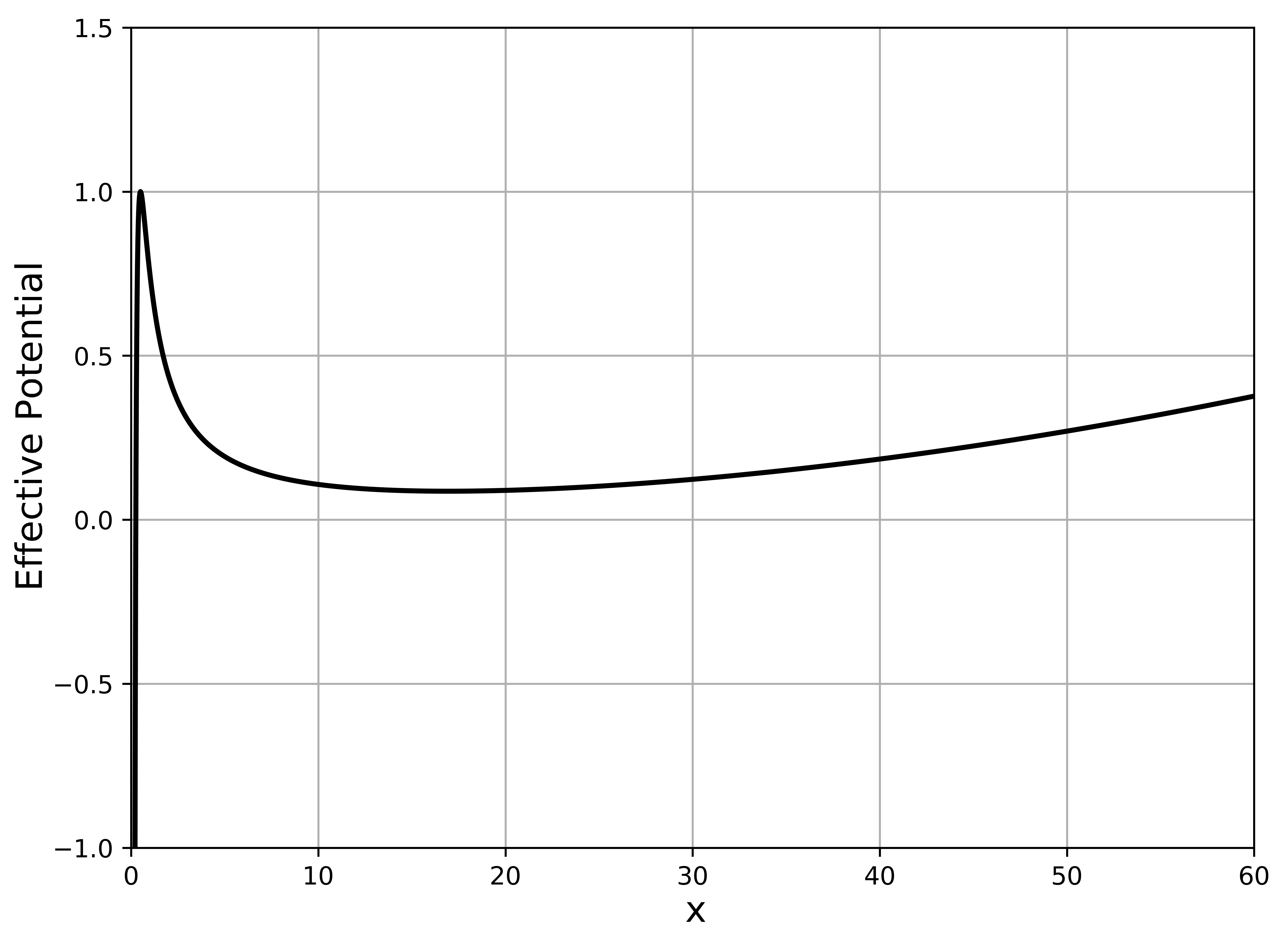}}
\caption{Effective Potential given by equation~(\ref{qd}).}
\label{fig7}
\end{figure}

\begin{figure}[!htb]
\centerline{\includegraphics[width=9.0cm]{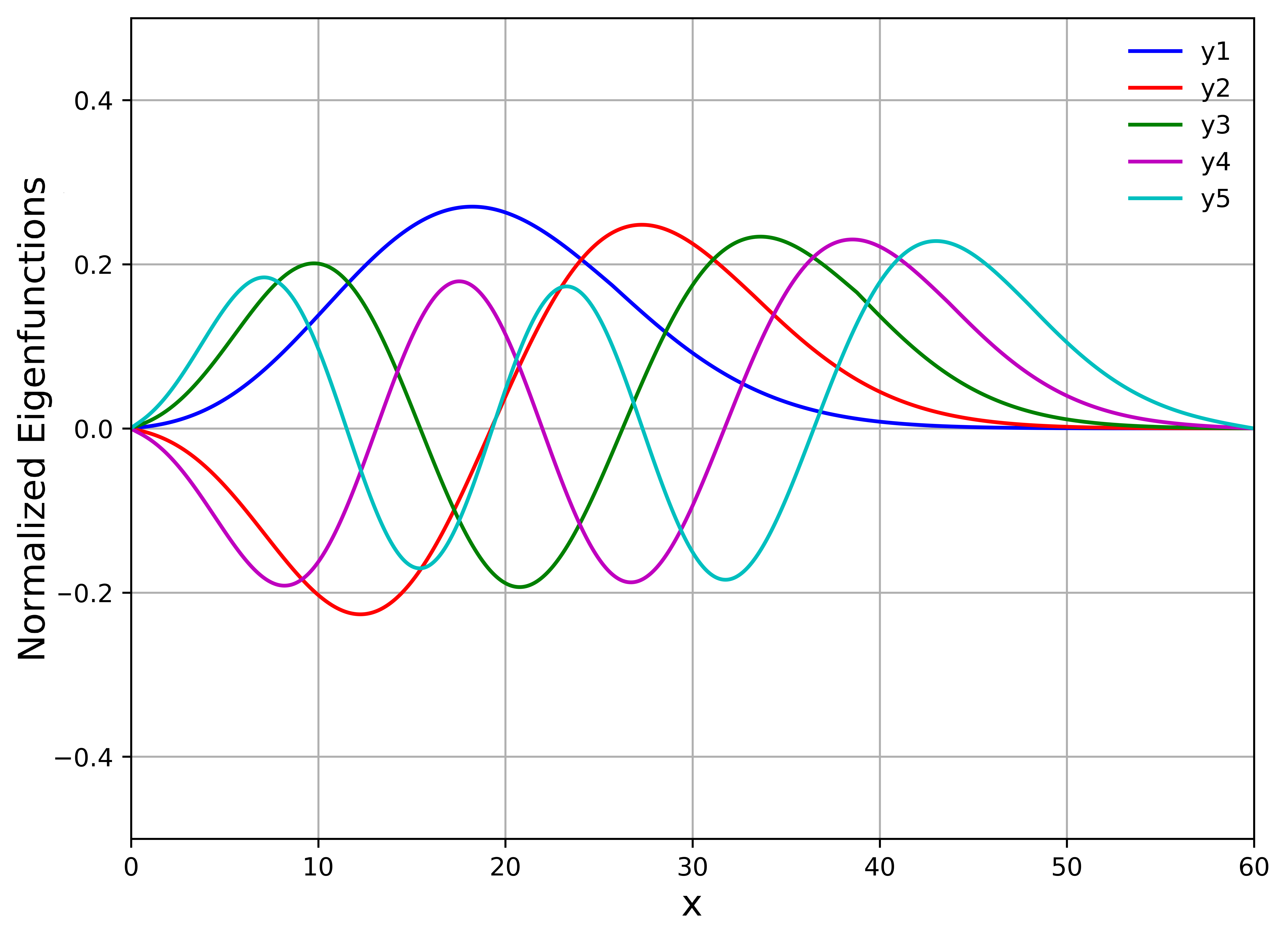}}
\caption{Normalized Eigenfunctions of equation~(\ref{qd}), with the potential given by equation~(\ref{morse}) with $\ell=0$.}
\label{fig8}
\end{figure}

Thus, adjusting the rest of the code, we obtain the first five eigenvalues. The effective potential and the eigenfunctions can be seen in Figures~\ref{fig7} and \ref{fig8}.


In Table \ref{tabela3}, we can compare the eigenvalues found by the Numerov's method with the analytical values for the quantum dot given by $\eta_{n\ell} = 2(n+\ell+1)\omega$.
\renewcommand{\arraystretch}{0.9}
\begin{table}[!ht]
  \caption{Comparison between the energy values for the Quantum dot, in Rydberg units, found by the Numerov's method and the analytical ones for $\ell=0$ and n=0 and 1.}\label{tabela3}
  \vspace*{0.2cm}
  \begin{center}
  \begin{tabular}{c|c|c}
    \hline
    \hline
      n           &Numerov's Energy    &Analytical value   \\ \hline
      4           & 0.1046             & 0.10  \\
      6           & 0.1403             & 0.14  \\
      8           & 0.1760             & 0.18  \\
      10          & 0.2134             & 0.22  \\
      12          & 0.2507             & 0.26   \\
      \hline
      \hline
  \end{tabular}
  \end{center}
\end{table}
\renewcommand{\arraystretch}{1}

\newpage
\section{Conclusion}

Analyzing the results arranged in Tables~\ref{tabela1}, \ref{tabela2} and~\ref{tabela3} we can conclude that the method used here is able to reproduce the analytical results within small errors.

Therefore, it is evident that the numerical method of Numerov is a powerful tool, easy to use, which can help in the development of not only new knowledge in the area of programming,
but also the solution of Schr\"{o}dinger equations outside the usual results found in the examples of modern physics books. We hope that this short introduction to the code, which can be found
in full at \url{https://1drv.ms/u/s!Ai_Lqkgh1kiskp5_cfOtwCfqX-LpTw?e=srdmzS}, will open doors for students to create their own versions, increasingly improving the versatility of this tool.

\section*{Acknowledgment}

One of us (FS) was financed in part by the Coordena\c{c}\~{a}o de Aperfei\c{c}oamento de Pessoal de N\'{\i}vel Superior -- Brazil (CAPES), Finance Code 001.

\renewcommand\refname{References}

\end{document}